# Are there too many uncited articles? Zero inflated variants of the discretised lognormal and hooked power law distributions[1]

Mike Thelwall
Statistical Cybermetrics Research Group, University of Wolverhampton, Wulfruna Street, Wolverhampton WV1 1ST, UK. E-mail: m.thelwall@wlv.ac.uk

Although statistical models fit many citation data sets reasonably well with the best fitting models being the hooked power law and discretised lognormal distribution, the fits are rarely close. One possible reason is that there might be more uncited articles than would be predicted by any model if some articles are inherently uncitable. Using data from 23 different Scopus categories, this article tests the assumption that removing a proportion of uncited articles from a citation dataset allows statistical distributions to have much closer fits. It also introduces two new models, zero inflated discretised lognormal distribution and the zero inflated hooked power law distribution and algorithms to fit them. In all 23 cases, the zero inflated version of the discretised lognormal distribution was an improvement on the standard version and in 15 out of 23 cases the zero inflated version of the hooked power law was an improvement on the standard version. Without zero inflation the discretised lognormal models fit the data better than the hooked power law distribution 6 out of 23 times and with it, the discretised lognormal models fit the data better than the hooked power law distribution 9 out of 23 times. Apparently uncitable articles seem to occur due to the presence of academic-related magazines in Scopus categories. In conclusion, future citation analysis and research indicators should take into account uncitable articles, and the best fitting distribution for sets of citation counts from a single subject and year is either the zero inflated discretised lognormal or zero inflated hooked power law.

## 1. Introduction

Citation-based indicators are commonly used to help assess academic departments (Hicks, 2012; Wilsdon, Allen, Belfiore, et al., 2015) and scholarly journals (Schweizer, 2010) as well as in some digital libraries to help rank search results (Lawrence, Giles, & Bollacker, 1999; Mayr, & Walter, 2007). Citation analysis is still controversial, however, for general reasons as well as for specific criticisms of individual indicators (e.g., DORA, 2012; MacRoberts & MacRoberts, 1989). In order to construct the most suitable indicators and to evaluate the limitations of existing indicators, it is useful to know as much about citations as possible, such as why they are created, what their role is and the factors that influence their creation (Borgman & Furner, 2002; Moed, 2006; van Raan, 2005). Four ways to achieve this are to theorise about the role of citations in scholarly communication (Merton, 1973), to ask scholars why they cite (Brooks, 1985; Case & Higgins, 2000), to examine individual citations to ascertain their apparent purpose (Chubin & Moitra, 1975; Oppenheim & Renn, 1978) and to statistically analyse collections of articles to identify factors that associate with citation counts (Peters & van Raan, 1994; Zitt & Bassecoulard, 1998). For statistical analyses, it is important to identify the overall distribution of sets of citation counts so that appropriate regression (e.g., linear, negative binomial, or zero inflated; any transformations needed) or other techniques can be selected (Thelwall, 2016b). The most appropriate choice of indicator also depends upon the nature of citation distributions. For instance, impact factors for journals should be calculated with the geometric mean rather than the arithmetic mean because of the skewed nature of citation counts for journals (Thelwall & Fairclough, 2015; Zitt, 2012). The distribution of sets of citation counts is also needed to assess the precision (Thelwall, 2016a) and other properties of indicators generated from citation counts as well as to help identify when sets of citation counts are anomalous in some way. There have been many attempts to identify statistical distributions that are appropriate for sets of citation counts but there is no consensus about which is the best overall.

The most appropriate sets of articles to examine from the perspective of citation count distributions are sets of journal articles because journal articles are the primary scholarly record in most areas of scholarship (excluding the arts and humanities and some social sciences). These sets should also be homogeneous in the sense of being from the same subject and year (or at least a common citation window: Abramo, Cicero, & D'Angelo, 2011) because different fields attract citations at different rates and so should not be mixed, if possible, and citations accumulate over time. For a typical homogeneous set of articles, the citation counts are highly skewed because few articles generated highly cited and many remain uncited (Seglen, 1992). If articles with few citations are excluded, then sets of citation counts fit a power law or a discretised lognormal distribution quite well (Garanina & Romanovsky, 2015; Redner, 1998; van Raan, 2001), but if all articles are included then the power law is a very poor fit for almost all subjects (exception: physics) and the discretised lognormal distribution and hooked power law (described below) tend to be much better (Eom & Fortunato, 2011; Radicchi, Fortunato, & Castellano, 2008; Thelwall & Wilson, 2014). Of these two, the hooked power law seems to fit better for a majority of subjects, although the discretised lognormal fits better for a minority (Thelwall & Wilson, 2014). A discrete version of the power law, the Yule-Simon process, also fits citation data reasonably well (Brzezinski, 2015) but cannot cope with subjects that have modes (most common values) greater than 0. A range of other distributions have also been proposed but all have problems. Since citation data sets are a type of count (integer) data, a range of count data models have been tested. The negative binomial distribution (Hilbe, 2011) model is a count model for skewed data sets but does not fit citation counts as well as the other two models (Low, Wilson, & Thelwall, 2015). Stopped sum models (Neyman, 1939) seem to fit slightly better but have parameter estimation issues that make them impractical to use (Low, Wilson, & Thelwall, 2015).

Although the discretised lognormal and hooked power law seem to be the distributions that fit sets of citation counts best, at least in practice, neither are perfect fits for most Scopus subjects and have particular problems in estimating the number of uncited articles (Thelwall, 2016c). This article assesses whether the main problem is that there are too many uncited articles published in journals, in the sense that if uncited articles are given special treatment then the remaining articles fit citation distributions much better. This could occur, for example, if some articles are almost uncitable because they make a valid contribution but either close off an area of research, address a highly niche topic, or are of interest only to students, practitioners or policy makers. This issue is investigated with a purely quantitative approach by systematically removing uncited articles in order to ascertain whether statistical distributions fit better afterwards. In addition, this article introduces two new distributions, the Zero Inflated Discretised Lognormal (ZIDL) and the Zero Inflated Hooked Power law (ZIHP) to deal with this situation, as well as software to fit them.

## 2. The zero inflated discretised lognormal and hooked power law distributions

The hooked power law, also known as the shifted power law (Pennock, Flake, Lawrence, Glover, & Giles, 2002), has probability mass function $f(n) = A(B + n)^{-\alpha}$, where α and $B$ are the parameters of the distribution and $A$ is determined by the choice of α and $B$ because the sum of f(x) for all theoretically possible values of $n$ must be 1. The parameter α is also found in the power law and primarily determines how high the citation counts are likely to be, and the shift parameter $B$ primarily affects the extent to which very low values occur, including the value of the mode.

The continuous lognormal distribution $ln\mathcal{N}(\mu, \sigma^2)$ has probability density function $f(x) = \frac{1}{x\sigma\sqrt{2\pi}}e^{-\frac{(\ln(x)-\mu)^2}{2\sigma^2}}$ with scale parameter μ and location parameter σ that are the mean and standard deviations of the natural log of the data (Limpert, Stahel, & Abbt, 2001). This can be converted into the discretised lognormal distribution $ln\ddot{\mathcal{N}}(\mu, \sigma^2)$ by integrating a unit interval around each positive integer, giving probability mass function $f(n) = \frac{1}{A}\int_{n-0.5}^{n+0.5} f(x)dx$, where $A = \int_{0.5}^{\infty} f(x)dx$ compensates for the missing interval (0,0.5] from the continuous distribution.

The discretised lognormal distribution is only defined for positive integers and not zero, so cannot accommodate uncited articles. The standard way to circumvent this issue, and the one used in this article, is to add 1 to all citation counts before analysing them. This is not necessary for the hooked power law but, if done, makes no difference except for decreasing the value of $B$ by 1.

## 3. Zero inflation calculations

A zero inflated variant of a distribution incorporates a procedure to remove some zeros from a dataset before applying the main distribution to the remaining data. This can be theorised in terms of some of the zeros being "natural" or predetermined (e.g., inherently uncitable articles) whereas the remainder are not. Given any dataset, it is typically not known whether any particular zero is predetermined and so it is necessary to estimate the proportion $p$ of the data are predetermined zeros. For a set of articles from a single subject and year, it would therefore be necessary to estimate the proportion that are inherently uncitable. After removing this estimated proportion of zeros from the data, the model can be fitted to the remaining data as normal. The probabilities of each value (i.e., citation count for an article) can then be calculated from the fitted model using the model parameters and $p$, as follows. For compatibility with the discretised lognormal distribution, the description assumes that 1 has been added to the data so that the zeros are in fact ones.

Suppose that $f(n)$ is a discrete probability density function defined for positive integers 1,2,... so that $\sum_{n=1}^{\infty} f(n) = 1$. As described above, a zero inflated model applied to a data set is simply a probability $p$ that a data value does not fit the model described by $f(n)$, in which case the data value is fixed at 1. Denote the zero inflated extension of $f(n)$ with probability $p$ by $f(n, p)$. If both $p$ and $f(n)$ have been estimated then the probability of any number greater $n$ than 1 is $(1 - p)f(n)$ because first the possibility of 1 must be rejected, with probability $(1 - p)$ and then $n$ must be selected from the main distribution, with probability $f(n)$. In contrast a 1 could occur either as a natural 1 with probability $p$ or, the 1 could occur from the model, with probability $f(1)$, after selecting the model with probability $(1 - p)$.

$$f(n,p) = \begin{cases} (1-p)f(n) & if \; n > 1 \\ p + (1-p)f(1) & if \; n = 1 \end{cases}$$

### *Fitting zero inflated distributions*

A statistical distribution is typically fitted to a data set by using an algorithm to select the free parameters of the distribution (i.e., the parameters with values that are allowed to vary) that maximise the probability that the data was derived from the model, known as maximum likelihood estimation. In practice, the natural logarithm of this probability is used, known as the log-likelihood. The log-likelihood of a dataset $\{x_i\}_{i=1}^N$ of size $N$ given a zero-inflated probability density function $f(n,p)$ is therefore $\ln(\prod f(x_i,p)) = \sum \ln(f(x_i,p))$.

Assume that a procedure is available to fit a discrete probability function $f(x)$ to a data set, such as the R package powerRlaw for the discretised lognormal distribution (Gillespie, 2013, 2015) or R code for the hooked power law (see below). Then this can be adapted to calculate the zero-inflated log-likelihood of the distribution as follows.

Let $N$ be the size of the dataset $\{x_i\}$ and let $r$ be the number of 1s in the dataset. For notational convenience, arrange the dataset so that these 1s are the first data points. For any $k \in \{0, \dots r\}$, remove $k$ of the 1s of the dataset $\{x_i\}$ to form a new dataset $\{x_i\}_{i=k+1}^N$ of size $N - k$. This dataset has $r - k$ ones. The maximum likelihood value of $p$ that will cause $k$ ones to be removed is $p = k/N$ since the size of the dataset is $N$. Using the probability density function $f_k(n)$ for the main distribution fitted to the truncated dataset $\{x_i\}_{i=k+1}^N$ (fitted with the standard non-zero inflated procedure), the zero truncated probability density function is therefore:

$$f_k(n,k/N) = \begin{cases} (1-k/N)f_k(n) & if \; n > 1 \\ k/N + (1-k/N)f_k(1) & if \; n = 1 \end{cases}$$

The log-likelihood for $f_k(n,k/N)$ for $k \in \{1, \dots r+1\}$ can then be derived from the log-likelihood of the main distribution $f_k(n)$ by adjusting the values greater than 1 for the extra parameter $p$, removing the component of the log-likelihood for the 1s in the fitted main model, and adding the probability of a 1 for all 1s in the dataset, as follows.

The log-likelihood of the complete data set for $k$ is given by

$\sum_{i=1}^N \ln(f_k(x_i,p)) = \sum_{i=1}^k \ln(f_k(x_i,p)) + \sum_{i=k+1}^N \ln(f_k(x_i,p))$

$= \sum_{i=1}^k \ln(\frac{k}{N} + (1-\frac{k}{N})f_k(1)) + \sum_{i=k+1}^N \ln((1-\frac{k}{N})f_k(x_i))$

$= k\ln(k/N + (1-\frac{k}{N})f_k(1)) + \sum_{i=k+1}^N \ln((1-\frac{k}{N})f_k(x_i))$

$= k\ln(k/N + (1-\frac{k}{N})f_k(1)) + \sum_{i=k+1}^N \ln(1-\frac{k}{N}) + \sum_{i=k+1}^N \ln f_k(x_i)$

$= k\ln(k/N + (1-\frac{k}{N})f_k(1)) + (N-k)\ln(1-\frac{k}{N}) + \sum_{i=k+1}^N \ln f_k(x_i)$

Therefore, to convert a log-likelihood for $f_k(n)$ for the truncated dataset $\{x_i\}_{i=k+1}^N$ into a log-likelihood for the zero inflated model $f_k(n,k/N)$ on the full dataset $\{x_i\}_{i=1}^N$, carry out the following:

- Subtract all the calculations associated with the $r - k$ ones from the original model. In logarithm terms, subtract $(r - k)\ln(f_k(1))$.
- Multiply the probabilities for the remaining by $1 - p$ because they were not selected for zero truncation. In logarithm terms, add $(N - r)ln(1 - \frac{k}{N})$.

- Add the values for the $r$ ones for the full dataset. In logarithm terms, add $r\ln(\frac{k}{N} + \left(1 - \frac{k}{N}\right)f_k(1))$.

With the above log-likelihood calculation procedure for zero-inflated variants of the discretised lognormal and the hooked power law, a procedure is needed to identify the parameters that maximise the log-likelihood of the data.

The exhaustive search method calculates the log-likelihood of all possible values of $p = k/N$ and selects the largest log-likelihood. This is inefficient and may be impractical for very large datasets but is guaranteed to find the correct solution in all cases, at least to an accuracy of $k/N$ in terms of $p$. This assumes that the behaviour of the log-likelihood function is well-behaved in the sense that small changes in p would produce small log-likelihood changes. This should be true for datasets that are not too small and are fitted by the main distribution reasonably well. If these assumptions fail then the fitted distribution would be sub-optimal. Moreover, the method ignores the possibility of negative $p$ values, even though these would be plausible given a situation in which some articles were artificially cited. A more precise approach to identify the parameters would be to use maximum likelihood estimation directly. A method has been developed that can achieve this (Smolinsky, 2016).

## 4. Research questions

The research questions ask whether the zero inflated versions of two distributions fit citation count data better than the standard versions. A positive answer would also support, but not prove, the hypothesis that there were uncitable articles within Scopus categories. In the case of the discretised lognormal distribution, this is subject to 1 being added to all citation counts first.

1. Does the zero inflated discretised lognormal distribution fit sets of citation counts for journal articles from a single subject and year better than the discretised lognormal distribution?
2. Does the zero inflated hooked power law distribution fit sets of citation counts for journal articles from a single subject and year better than the hooked power law?
3. Which out of the four distributions named above tends to be the best fitting for sets of citation counts for journal articles from a single subject and year?

## 5. Data and methods

The data was re-used from a previous paper (Thelwall, 2016c). It consists of citation counts for journal articles published in 2006 (excluding non-article documents, such as reviews) from 23 Scopus categories with up to 10,000 citation counts per category. The citation counts to date were downloaded from Scopus in November 2015. The year 2006 gives a long enough time for citation counts to mature and for time factors within the year (e.g., January or December publication) to be relatively unimportant. Only the first 5000 and last 5000 articles published in 2006 were included for categories with more than 10000 articles, a technical limitation because Scopus reports a maximum of 5000 results per query and allows sorting by date and reverse date order. The 23 categories (see tables for category names) represent a systematic wide selection of relatively narrow subject areas. Broad subject areas were avoided because these could mix different citation distributions and hence give less clear results.

As described above, 1 was added to all citation counts (for both cited and uncited articles) in order that the discretised lognormal distribution could be fitted to complete data sets, using the exhaustive method. After this, the standard and zero-inflated variants of the discretised lognormal and hooked power law were fitted to each of the 23 data sets in order to compare how well they fit. For both distributions, the main model was fitted using maximum likelihood estimation. Although there are different methods to test whether one distribution fits a data set better than another a standard technique is to select the one that gives the lowest Akaike information criterion (AIC) value (Akaike, 1974). This is based on the log-likelihood of the distributions but compensating for differences in the numbers of parameters. This penalises the log-likelihood of a distribution by 1 for each additional parameter than it has. Since zero-inflated variants of distributions have one extra parameter, $p$, their log-likelihoods need to be greater than 1 larger than the log-likelihoods of the standard models in order to be selected as better. Although AIC is suitable for comparing nested models (Akaike, 1974; Burnham & Anderson, 2002, p. 81), as is the case here, a disadvantage of the AIC test is that it does not give a probability that one distribution fits better than another and so it cannot be used for hypothesis testing. Although the Vuong test (Vuong, 1989) is commonly used to fill this gap, it is not applicable in the case of zero inflation because it can give misleading results (Wilson, 2015).

Although it is rarely used to compare distributions, a commonly used method to assess an individual distribution is the Kolmogorov-Smirnov goodness of fit test (Massey, 1951; Pettitt & Stephens, 1977). This reports the maximum degree to which the cumulative distribution function for a fitted model differs from the empirical cumulative distribution function for the data. Smaller values indicate that the theoretical distribution tends to be a better fit. This was used as a second method to assess the fits of the zero inflated distributions.

Finally, the fit of the zero inflated and standard distributions were compared visually by constructing videos of graphs of the empirical and theoretical cumulative distributions in order to assess intuitively how increasing the zero inflation parameter $p$ affects the overall fit of the data. Although Q-Q plots are generally preferred to comparing cumulative distributions, they can be unhelpful for discrete distributions and can be misleading for the lognormal distribution (Das, & Resnick, 2008) and so were not used.

Instructions for accessing the code used in the methods are in the Appendix.

# 6. Results

## *Zero inflated discretised lognormal distribution*

In all cases the log-likelihood for the zero inflated variant is at least 1 higher than for the standard variant, indicating a better model (Table 1). In all cases too, the Kolmogorov-Smirnov statistic is lower for the zero inflated variant, indicating a better fit. Even without an equivalent to the Vuong statistic to conduct a hypothesis test, the fact that an improvement is universal is strong evidence that the zero inflated variants are an improvement for the discretised lognormal distribution.

**Table 1**. The results of fitting Discretised lognormal results. Uncited is the number of uncited articles included in the main model and uncitable is the percentage of articles without citations removed before fitting the main model. The top row for each subject is the non-zero inflated variant without any of the uncited articles classed as uncitable and the bottom row is the zero inflated variant (i.e., with the number of uncited articles removed determined by the algorithm described above).

| Subject | Articles | Uncitable | Uncited | μ | σ | K-S | Log-lik. |
|---|---|---|---|---|---|---|---|
| Food Science | 9992 | 0% | 789 | 2.54 | 1.26 | 0.059 | -41863.0 |
| Food Science | 9992 | 7% | 128 | 2.74 | 1.07 | 0.029 | -41509.4 |
| Cancer Research | 9994 | 0% | 847 | 2.77 | 1.40 | 0.067 | -45194.0 |
| Cancer Research | 9994 | 7% | 104 | 3.02 | 1.16 | 0.029 | -44716.2 |
| Marketing | 2260 | 0% | 179 | 2.43 | 1.33 | 0.033 | -9327.56 |
| Marketing | 2260 | 5% | 68 | 2.58 | 1.20 | 0.015 | -9301.14 |
| Filtration and Separation | 3282 | 0% | 359 | 2.18 | 1.39 | 0.048 | -12879.3 |
| Filtration and Separation | 3282 | 6% | 153 | 2.38 | 1.24 | 0.039 | -12842.8 |
| Physical and Theoretical Chemistry | 9986 | 0% | 584 | 2.46 | 1.18 | 0.032 | -40335.4 |
| Physical and Theoretical Chemistry | 9986 | 4% | 205 | 2.57 | 1.08 | 0.015 | -40221.5 |
| Computer Science Applications | 8148 | 0% | 888 | 2.19 | 1.40 | 0.043 | -32176.9 |
| Computer Science Applications | 8148 | 6% | 382 | 2.39 | 1.25 | 0.028 | -32089.0 |
| Management Sci. & Operations Research | 3993 | 0% | 249 | 2.45 | 1.24 | 0.030 | -16300.3 |
| Management Sci. & Operations Research | 3993 | 3% | 112 | 2.55 | 1.15 | 0.015 | -16270.9 |
| Geochemistry and Petrology | 8292 | 0% | 268 | 2.79 | 1.12 | 0.034 | -35838.5 |
| Geochemistry and Petrology | 8292 | 2% | 72 | 2.86 | 1.04 | 0.021 | -35744.1 |
| Economics and Econometrics | 9974 | 0% | 1004 | 2.23 | 1.39 | 0.024 | -39594.6 |
| Economics and Econometrics | 9974 | 5% | 475 | 2.39 | 1.26 | 0.014 | -39511.4 |
| Energy Engineering & Power Technology | 7833 | 0% | 1879 | 1.33 | 1.79 | 0.045 | -27361.9 |
| Energy Engineering & Power Technology | 7833 | 15% | 678 | 2.02 | 1.41 | 0.026 | -27204.2 |
| Computational Mechanics | 7776 | 0% | 564 | 2.19 | 1.17 | 0.026 | -29197.8 |
| Computational Mechanics | 7776 | 3% | 297 | 2.28 | 1.09 | 0.018 | -29158.2 |
| Global and Planetary Change | 834 | 0% | 36 | 3.00 | 1.35 | 0.048 | -3931.5 |
| Global and Planetary Change | 834 | 3% | 11 | 3.10 | 1.24 | 0.031 | -3921.9 |
| Virology | 6534 | 0% | 261 | 2.81 | 1.05 | 0.058 | -27942.5 |
| Virology | 6534 | 4% | 19 | 2.91 | 0.91 | 0.028 | -27672.4 |
| Metals and Alloys | 9964 | 0% | 2250 | 1.22 | 1.53 | 0.011 | -31292.5 |
| Metals and Alloys | 9964 | 5% | 1710 | 1.42 | 1.43 | 0.011 | -31277.9 |
| Control and Optimization | 1043 | 0% | 77 | 2.08 | 1.11 | 0.024 | -3742.11 |
| Control and Optimization | 1043 | 3% | 44 | 2.15 | 1.04 | 0.017 | -3737.9 |
| Critical Care & Intensive Care Medicine | 2625 | 0% | 514 | 1.75 | 1.82 | 0.065 | -10199.0 |
| Critical Care & Intensive Care Medicine | 2625 | 13% | 176 | 2.33 | 1.47 | 0.043 | -10141.6 |
| Developmental Neuroscience | 1394 | 0% | 47 | 2.83 | 1.08 | 0.038 | -6039.9 |
| Developmental Neuroscience | 1394 | 3% | 7 | 2.91 | 0.98 | 0.021 | -6009.51 |
| Pharmaceutical Science | 9228 | 0% | 1744 | 1.90 | 1.61 | 0.082 | -35373.1 |
| Pharmaceutical Science | 9228 | 16% | 293 | 2.48 | 1.20 | 0.048 | -34907.2 |
| Nuclear and High Energy Physics | 9994 | 0% | 984 | 2.34 | 1.41 | 0.042 | -40965.9 |
| Nuclear and High Energy Physics | 9994 | 6% | 371 | 2.54 | 1.25 | 0.022 | -40832.3 |
| Neuropsychology & Physiological Psych. | 2927 | 0% | 271 | 2.59 | 1.39 | 0.074 | -12706.8 |
| Neuropsychology & Physiological Psych. | 2927 | 8% | 43 | 2.85 | 1.16 | 0.044 | -12591.8 |
| Health social science | 4352 | 0% | 704 | 2.15 | 1.50 | 0.083 | -17336.6 |
| Health social science | 4352 | 14% | 76 | 2.64 | 1.11 | 0.040 | -17036.3 |
| Cultural Studies | 4848 | 0% | 2137 | -0.38 | 1.73 | 0.010 | -10601.4 |
| Cultural Studies | 4848 | 12% | 1570 | 0.30 | 1.50 | 0.004 | -10594.3 |
| Health Information Management | 697 | 0% | 102 | 1.96 | 1.37 | 0.046 | -2569.65 |
| Health Information Management | 697 | 11% | 27 | 2.28 | 1.12 | 0.033 | -2550.74 |

Visual inspections of the differences between the empirical and theoretical cumulative distribution functions for the main model confirm that the zero inflation does allow it to fit more closely to the data, although there are still problems with either the top or bottom of the distribution, indicating a remaining systematic problem. This is illustrated in Figures 1 (not-zero inflated) and 2 (Zero inflated) for Pharmaceutical Science and videos are available online for all possible zero inflated versions of each main model (see Appendix).

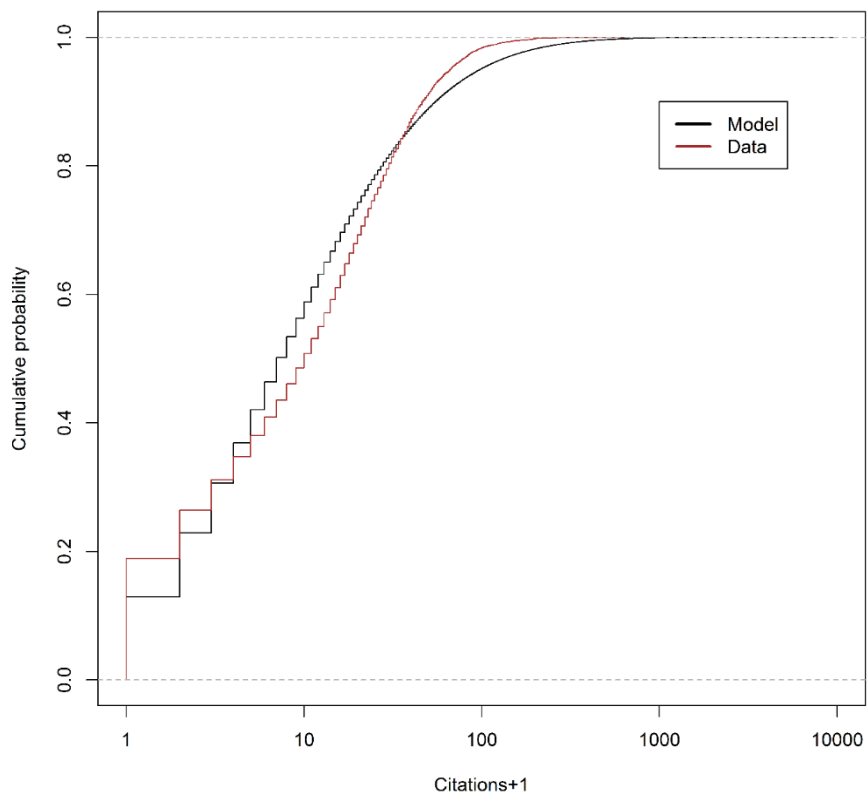

**Figure 1**. The cumulative distribution for the standard (not zero-inflated) discretised lognormal distribution and the data for Pharmaceutical Science.

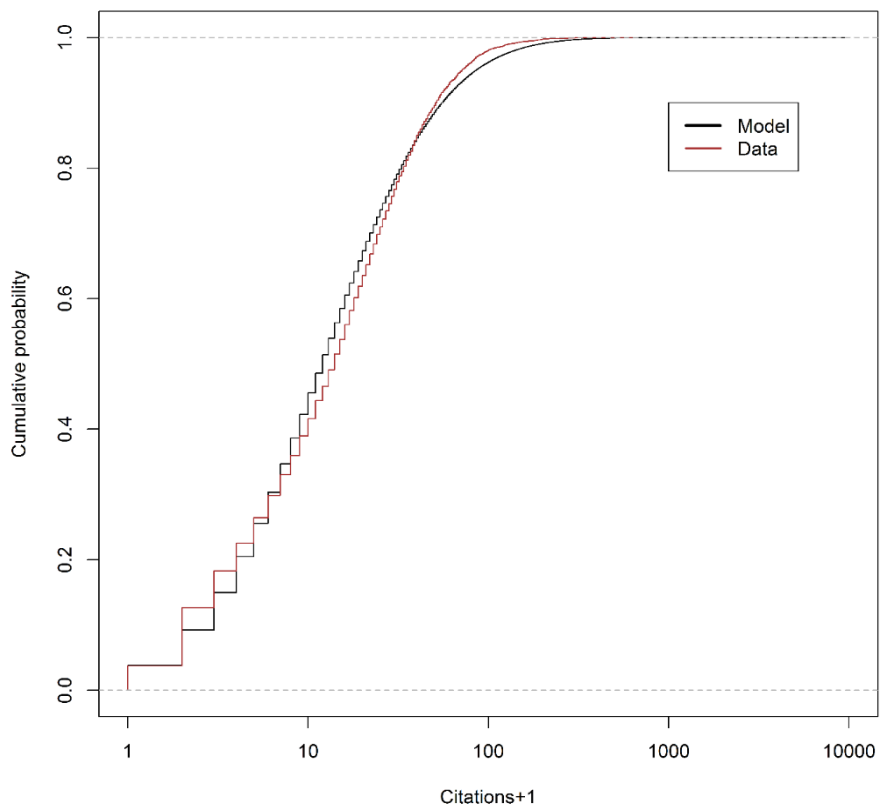

**Figure 2**. The cumulative distribution for the zero-inflated discretised lognormal distribution and the data for Pharmaceutical Science.

## *Zero inflated hooked power law distribution*

In 7 cases (in bold below), zero inflation does not increase the log-likelihood and in 1 case (in italic below) zero inflation does not increase the log-likelihood enough (1) to recommend the model via AIC. In the remaining 15 cases the log-likelihood for the zero inflated variant is at least 1 higher than for the standard variant, indicating a better model. In all these 15 cases the Kolmogorov-Smirnov statistic is lower for the zero inflated variant, indicating a better fit.

**Table 2**. Hooked power law results. Uncited is the number of uncited articles included in the main model and uncitable is the percentage of articles without citations removed before fitting the main model. The top row for each subject is the non-zero inflated variant without any of the uncited articles classed as uncitable and the bottom row is the zero inflated variant (i.e., with the number of uncited articles removed determined by the algorithm described above).

| Subject* | Articles | Uncitable | Uncited | α | B | K-S | Log-lik. |
|---|---|---|---|---|---|---|---|
| Food Science | 9992 | 0% | 789 | 5.76 | 89.8 | 0.028 | -41510.1 |
| Food Science | 9992 | 3% | 458 | 6.69 | 115.8 | 0.014 | -41430.1 |
| Cancer Research | 9994 | 0% | 847 | 3.94 | 67.9 | 0.043 | -44859.9 |
| Cancer Research | 9994 | 5% | 352 | 4.55 | 93.1 | 0.019 | -44652.6 |
| Marketing | 2260 | 0% | 179 | 3.52 | 37.8 | 0.016 | -9304.07 |
| Marketing | 2260 | 2% | 132 | 3.72 | 43.2 | 0.008 | -9298.32 |
| Filtration and Separation | 3282 | 0% | 359 | 3.56 | 31.5 | 0.045 | -12851.5 |
| Filtration and Separation | 3282 | 5% | 208 | 4.22 | 45.8 | 0.044 | -12820.1 |
| **Physical and Theoretical Chemistry** | **9986** | **0%** | **584** | **4.77** | **59.3** | **0.02** | **-40235.6** |
| **Physical and Theoretical Chemistry** | **9986** | **0%** | **584** | **4.77** | **59.3** | **0.02** | **-40235.6** |
| Computer Science Applications | 8148 | 0% | 888 | 3.11 | 24.3 | 0.033 | -32110.0 |
| Computer Science Applications | 8148 | 4% | 563 | 3.4 | 31.1 | 0.015 | -32053.6 |
| **Management Sci. & Operations Research** | **3993** | **0%** | **249** | **4.08** | **47.6** | **0.008** | **-16261.3** |
| **Management Sci. & Operations Research** | **3993** | **0%** | **245** | **4.09** | **47.9** | **0.008** | **-16261.2** |
| **Geochemistry and Petrology** | **8292** | **0%** | **268** | **6.55** | **126.9** | **0.037** | **-35768.8** |
| **Geochemistry and Petrology** | **8292** | **0%** | **268** | **6.55** | **126.9** | **0.037** | **-35768.8** |
| Economics and Econometrics | 9974 | 0% | 1004 | 3.08 | 24.4 | 0.022 | -39548.2 |
| Economics and Econometrics | 9974 | 3% | 697 | 3.3 | 29.7 | 0.011 | -39506.9 |
| Energy Engineering & Power Technology | 7833 | 0% | 1879 | 2.07 | 4.7 | 0.069 | -27555.3 |
| Energy Engineering & Power Technology | 7833 | 16% | 647 | 2.83 | 16.7 | 0.040 | -27233.2 |
| **Computational Mechanics** | **7776** | **0%** | **564** | **4.87** | **46.1** | **0.008** | **-29142.3** |
| **Computational Mechanics** | **7776** | **0%** | **564** | **4.87** | **46.1** | **0.008** | **-29142.3** |
| Global and Planetary Change | 834 | *0%* | 36 | 3.5 | 67 | 0.020 | -3918.66 |
| Global and Planetary Change | 834 | *1%* | 29 | 3.56 | 70.3 | 0.020 | -3918.02 |
| **Virology** | **6534** | **0%** | **261** | **14.74** | **329.5** | **0.058** | **-27779.2** |
| **Virology** | **6534** | **0%** | **261** | **14.74** | **329.5** | **0.058** | **-27779.2** |
| Metals and Alloys | 9964 | 0% | 2250 | 2.38 | 5.2 | 0.026 | -31368.4 |
| Metals and Alloys | 9964 | 7% | 1552 | 2.62 | 7.8 | 0.017 | -31317.6 |
| **Control and Optimization** | **1043** | **0%** | **77** | **5.07** | **41.9** | **0.028** | **-3742.21** |
| **Control and Optimization** | **1043** | **0%** | **77** | **5.07** | **41.9** | **0.028** | **-3742.21** |
| Critical Care & Intensive Care Medicine | 2625 | 0% | 514 | 2.06 | 7.1 | 0.074 | -10272.2 |
| Critical Care & Intensive Care Medicine | 2625 | 13% | 161 | 2.78 | 23.2 | 0.054 | -10147.8 |
| **Developmental Neuroscience** | **1394** | **0%** | **47** | **11.34** | **258.1** | **0.041** | **-6016.32** |
| **Developmental Neuroscience** | **1394** | **0%** | **47** | **11.34** | **258.1** | **0.041** | **-6016.32** |
| Pharmaceutical Science | 9228 | 0% | 1744 | 2.95 | 19.4 | 0.098 | -35429.3 |
| Pharmaceutical Science | 9228 | 14% | 464 | 5.31 | 70.4 | 0.037 | -34768.1 |
| Nuclear and High Energy Physics | 9994 | 0% | 984 | 3.23 | 30.8 | 0.032 | -40878.1 |
| Nuclear and High Energy Physics | 9994 | 4% | 580 | 3.6 | 40.6 | 0.017 | -40793.1 |
| Neuropsychology & Physiological Psych. | 2927 | 0% | 271 | 4.55 | 72.6 | 0.051 | -12603.3 |
| Neuropsychology & Physiological Psych. | 2927 | 5% | 115 | 5.61 | 107.8 | 0.017 | -12541.1 |
| Health social science | 4352 | 0% | 704 | 3.82 | 37.9 | 0.092 | -17258.1 |
| Health social science | 4352 | 12% | 195 | 6.73 | 108.1 | 0.012 | -16976.4 |
| Cultural Studies | 4848 | 0% | 2137 | 2.26 | 1.3 | 0.021 | -10627.3 |
| Cultural Studies | 4848 | 18% | 1282 | 2.62 | 3.2 | 0.007 | -10598.5 |
| Health Information Management | 697 | 0% | 102 | 3.5 | 23.8 | 0.050 | -2561.02 |
| Health Information Management | 697 | 7% | 52 | 4.24 | 37.4 | 0.022 | -2548.68 |

*Bold: zero inflation does not increase the log-likelihood; italic below; zero inflation does not increase the log-likelihood enough (1) to pass the AIC test.

Visual inspections of the differences between the empirical and theoretical cumulative distribution functions for the main model again confirm that the zero inflation does allow it to fit more closely to the data, although there are again still problems with either the top or bottom of the distribution, indicating a remaining systematic problem. The improved fit is illustrated in Figures 3 (not-zero inflated) and 4 (zero inflated) for Pharmaceutical Science and videos are available online for all possible zero inflated versions of each main model (see Appendix).

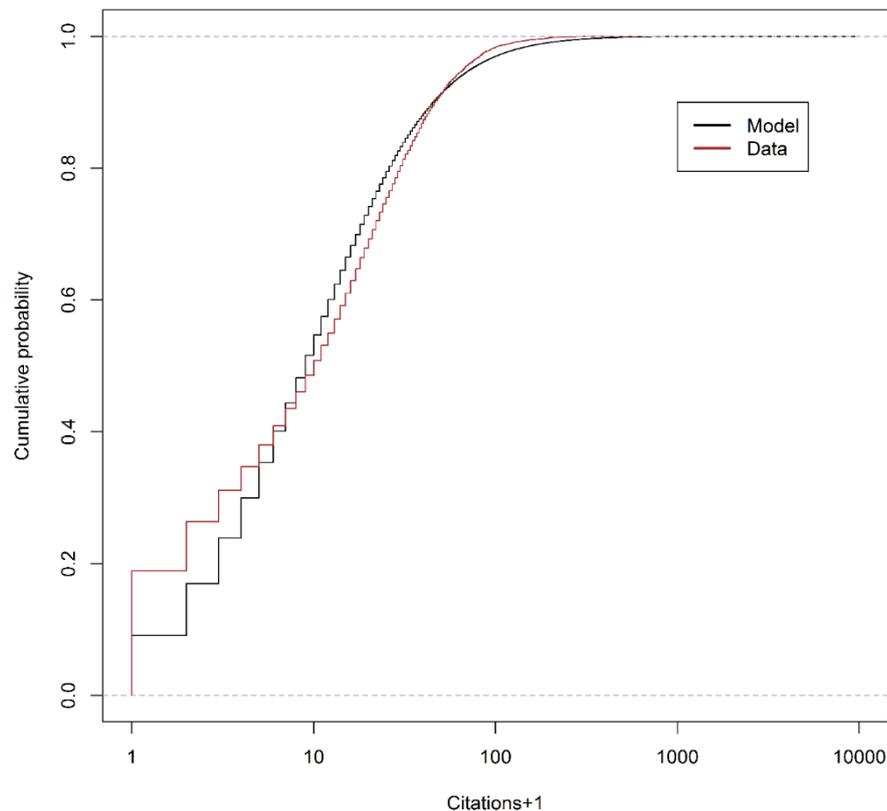

**Figure 3**. The cumulative distribution for the standard (not zero-inflated) hooked power law distribution and the data for Pharmaceutical Science.

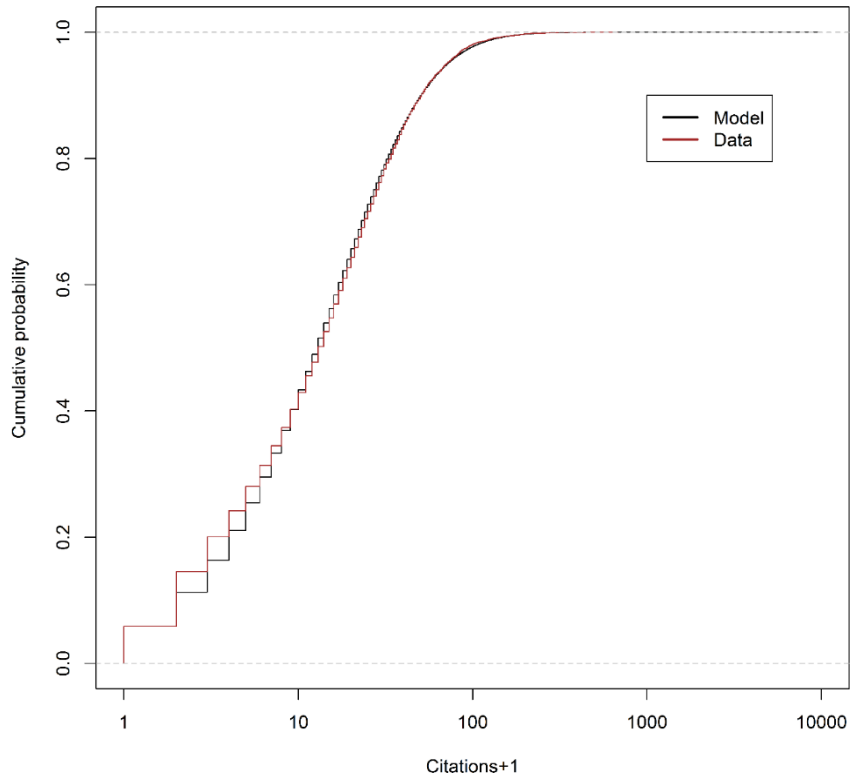

**Figure 4**. The cumulative distribution for the zero-inflated hooked power law distribution and the data for Pharmaceutical Science.

### Best fitting distribution overall

Without zero inflation, the best fitting distribution is the hooked power law in 17 out of 23 cases, in terms of the highest log-likelihood (Tables 1,2). With zero inflation (excluding it when it does not improve), the best fitting distribution is the hooked power law 14 out of 23 times. Overall, the hooked power law still tends to be a better fit, although the gap is narrower. In only two cases (Management Science and Operations Research; Computational Mechanics) the best fitting model does not use zero inflation, giving strong evidence for the zero inflation strategy overall.

## 7. Discussion

An important limitation for the results is that the Scopus categories are likely to contain a mix of sub-fields with different citation norms, such as qualitative and quantitative research that addresses the same topic. In addition, interdisciplinary research has citation properties that are different from each of its constituent fields (Levitt & Thelwall, 2008; Rinia, Van Leeuwen, & Van Raan, 2002). Such articles can create anomalies in the citation distributions of whichever subject area they are categorised within. The citation counts of the individual articles within each category may also be affected by other factors, such as the nationality of the authors, the number of co-authors, the length of the paper and even the readability of the abstract (Didegah & Thelwall, 2013; Gazni, 2011), all of which may cause anomalies when such articles are aggregated together within a category. In theory, this problem could be resolved by modelling citation counts with regression using all potential influences as independent variables. In such a case, the regression residuals could be expected to form a much purer distribution, with many of the anomalies removed.

Given the improvements in model fit gained by zero inflation for most subject areas, it is possible that the better fits reported for stopped sum models (Low, Wilson, & Thelwall, 2015) may be partly due to their higher predictions of zeros. It is not clear whether the relatively high goodness of fit of stopped sum models is because they are genuinely better models of the citation process, however, or essentially as an accidental by-product of their higher levels of prediction of zeros (if true). Future work may be able to compare stopped sum models and zero inflated discretised lognormal and hooked power law distributions to see which tends to fit best overall and to examine in detail how stopped sum models deal with zeros.

## Uncitable articles

The improved fit of zero inflated distributions in almost all cases (21 out of 23 categories are best fit by a zero inflated model) suggests that there may be a problem with the number of uncited articles. This statistical result does not prove that uncitable articles genuinely exist because the distributions tested might be incorrect in ways that cause them to underestimate the number of zeros. For example, another statistical model might fit the data better and not have a problem with excess zeros. With the zero inflated versions of the discretised lognormal distribution, the proportion of documents that are estimated to be uncitable varied from 2% (Geochemistry and Petrology) to 16% (Pharmaceutical Science) of all articles in the category and year. When the zero inflated version of the hooked power law is fitted then the proportion of documents that are estimated to be uncitable varied from 0% (many) to 18% (Cultural Studies) of all articles in the category and year.

Pharmaceutical Science is an extreme case in that the best fitting solution estimates that 83% of the uncited articles are apparently uncitable (i.e., natural zeros) with the discretised lognormal distribution, although this proportion is lower for the hooked power law. An investigation of the articles from this category found that some large indexed publications had a very high proportion of uncited articles. These seemed to be news-related industry-focused magazines. Although these may well be of interest to academics, they seem out of place in the sense of high volume publishing combined with few citations. The magazines included Deutsche Apotheker Zeitung (117 articles, 110 of which were uncited), Pharmaceutisch Weekblad (259 articles, 238 uncited), and Pharmazeutische Industrie (171 articles, 129 uncited). These are all German but there were also smaller similar journals from elsewhere, including Farmaceutisch Tijdschrift voor Belgie (8 articles, 7 uncited) and Atencion Farmaceutica (26 articles, 19 uncited). In contrast (and surprisingly), all of the 1263 Bioorganic and Medicinal Chemistry Letters articles from 2006 in Scopus had been cited (this was cross-checked and confirmed in the Web interface to Scopus: only erratum and Corrigendum were uncited in this journal in 2006). Some of the articles in the rarely cited publications were clearly magazine articles rather than scholarly research, such as "Franchising in drugstores is instructive for pharmacists" and "Drugstores keep coming" in Pharmaceutisch Weekblad. This suggests that at least some of the Scopus categories include non-academic papers that are classified as articles, affecting the overall citation distributions of the categories. Thus the problem of uncitable articles seems to be genuine.

Scopus acknowledges that it indexes "Over 360 trade publications" (https://www.elsevier.com/solutions/scopus/content, 16 March 2016) but does not seem to explicitly record them as such within the system in a way that would allow their content to be filtered out. For example, in March 2016 Deutsche Apotheker Zeitung had 18,954 documents indexed in Scopus, and 2984 were classified as "article", which is the same classification as given to journal articles. Moreover, the Scopus information page for this magazine did not give it a typology so that its status as a trade publication could not be checked. The same seems to be true for the Web of Science. For example, Pharmazeutische Industrie has 2233 documents indexed in the Web of Science, 2010 of which are classified as articles and the journal information page does not mention that it might be a trade publication. Scopus does in fact have a source type classification and a member of Elsevier has shared it with the author. In this classification spreadsheet, some sources are classified as Trade Journal, although Deutsche Apotheker Zeitung is classified as an academic journal. A journal (not in the data used here) that is classified as a trade journal in the Scopus spreadsheet is Telecommunication Journal of Australia, but this is not evident from the journal information page or the document classifications within Scopus (498 articles). Overall, then, a possible solution to the issue of trade publications would be to develop a manually curated list of trade publications to exclude, perhaps starting with official lists from Elsevier or Thomson Reuters.

Cultural Studies is another example with a high proportion of apparently uncitable articles. In this case the underlying cause seemed to be the inclusion of a number of intellectual magazines that may contain materials of interest to Cultural Studies scholars but that are rarely cited. Although only 3% of the articles in the category were published in journals that received no citations, 12% of the Cultural Studies articles were published in journals that contained at least 90% uncited articles, in contrast to 11% of articles that were published in journals containing 100% cited articles, such as the Journal of Ethnobiology and Ethnomedicine (47 articles, all cited). Examples of the intellectual magazines include North American Review ("A national treasure of thought and discussion": 41 articles, all uncited), Sind und Form ("Contributions to literature edited by the Academy of Arts": 50 articles, all uncited) and Merkur ("A German magazine for European thought": 109 articles, 103 uncited).

Some of the magazines discussed above have a few cited articles and it seems likely that most occasionally publish cited articles. Hence, it is an oversimplification to describe their articles as uncitable. Instead, they could be fairly characterised as extremely unlikely to be cited. The presence of magazines within a category may therefore also inflate the low citation counts in addition to the zeros. This may prevent any distribution from fitting well. Nevertheless, it seems reasonable to conclude that the problem of zero inflation is genuine and is due to the publication of substantial numbers of essentially uncitable articles within a category. These seem to be typically derived from magazines that are of interest to academics but that do not produce primary scholarship. This issue is not unique to Scopus because the Web of Science also indexed *Pharmazeutische Industrie* until 2013 and currently indexes *North American Review*.

## 8. Conclusions

The results show that there is a genuine problem of zero inflation within a substantial fraction of Scopus categories. Moreover, the zero inflated variant of the discretised lognormal distribution always seems to fit better than the standard version and the zero inflated version of the hooked power law distribution usually fits better than the standard version (in terms of AIC comparisons). The zero inflated hooked power law fits most subjects better than the zero inflated discretised lognormal distribution, but the difference the number of categories in each case is not large. Overall, the best fitting distribution out of the four tested is zero inflated for 21 out of the 23 categories. The investigations above confirmed that zero inflation is necessary due to the presence of essentially uncitable (in practice, rarely cited) magazine articles within Scopus categories.

The implications of zero inflation for scientometrics are substantial. Field normalised citation indicators (Waltman, van Eck, van Leeuwen, Visser, & van Raan, 2011) and percentile-based indicators (Waltman & Schreiber, 2013) can be influenced by the presence of uncitable articles. This is because these indicators are calculated with reference to the world average and so anything that artificially reduces the world average can artificially inflate indictor values. If a research unit publishes a substantial proportion of its articles in categories with many uncitable articles then, assuming that it does not publish in the uncitable magazines, the extra zeros in the data will lower the world average citation count for articles for the category and hence inflate the field normalised citation score of the group because this is calculated relative to the world average. Similarly, if there are many uncitable articles in a category then this will lower the number of citations required to get into the nth percentile, for any value of n, and hence inflate the proportion of the group's publications in the top nth percentile. The same is true if the comparisons are between individual scholars or entire countries (Fairclough & Thelwall, 2015; King, 2004). For example, since there were many German language magazines within the Pharmaceutical Science category, indicators for German-speaking nations for pharmacy or that include a substantial proportion of pharmacy articles are likely to be reduced by the presence of their uncitable articles in these magazines. In contrast, pharmacy-related indicators for non-German speaking nations are likely to be inflated by the artificial lowering of the world average. The fundamental issue here of the need to include only comparable sources when constructing normalised indicators has already been explicitly acknowledged and solved for the Leiden Ranking (Waltman, Calero-Medina, Kosten, et al., 2012; see also: van Raan, Van Leeuwen, & Visser, 2011), which goes further by developing a concept of "core journals", which are not only genuine academic journals but are also in English, international, and systematically referencing other core journals (http://www.leidenranking.com/methodology/indicators#core-publications).

Future statistical modelling approaches for citation distributions also need to take into account the potential presence of uncitable articles when attempting to model collections of articles, even if the two zero inflated models introduced here are not used (see also: Smolinsky, 2016). The uncitable articles could be accommodated either by filtering (see below), by adopting a zero-inflated variant of a distribution, or by using a model that naturally allowed high numbers of uncited articles.

The presence of uncitable articles is also likely to affect regression analyses that attempt to determine the factors that influence citation counts, such as abstract readability, title length, author nationality and collaboration type (Didegah & Thelwall, 2013). For this, filtering out uncitable articles (see below) before regression (for recommendations see: Thelwall, 2016b) would be the preferable strategy since the uncitable magazine articles are presumably completely irrelevant. Alternatively, a zero inflated regression method could be used. Although zero inflated negative binomial regression methods are available already, these are not adequate models and a regression algorithm is needed for the zero inflated discretised lognormal distribution instead. This is preferable to the zero inflated hooked power law since the latter has less precise parameter estimates.

A potential generic solution to the zero inflation problem is pre-filtering of subject categories in an attempt to remove all or most of the uncitable articles. A simple method for this would be to set a cited articles threshold T for each journal so that its articles would all be removed unless at least T% of them had been cited. For example, setting T=90 would remove most magazines in each category. This might also remove low impact academic journals and might affect non-English language journals disproportionately. Overall, however, it may be a simple way to improve the quality of data used in scientometric studies. Future research is needed to assess the efficacy of this approach or to suggest alternative strategies.

## 9. References

Abramo, G., Cicero, T., & D'Angelo, C. A. (2011). Assessing the varying level of impact measurement accuracy as a function of the citation window length. Journal of Informetrics, 5(4), 659-667.

Akaike, H. (1974). A new look at the statistical model identification. IEEE Transactions on Automatic Control, 19(6), 716-723.

Borgman, C. L., & Furner, J. (2002). Scholarly Communication and Bibliometrics. Annual Review of Information Science and Technology (ARIST), 36, 3-72.

Brzezinski, M. (2015). Power laws in citation distributions: Evidence from Scopus. Scientometrics, 103(1), 213-228.

Brooks, T. A. (1985). Private acts and public objects: An investigation of citer motivations. Journal of the American Society for Information Science, 36(4), 223-229.

Burnham, K. P. & Anderson, D. R. (2002). Model selection and multimodel inference: A practical information-theoretic approach (2nd ed.), Berlin: Springer-Verlag.

Case, D. O., & Higgins, G. M. (2000). How can we investigate citation behavior? A study of reasons for citing literature in communication. Journal of the American Society for Information Science, 51(7), 635-645.

Chubin, D. E., & Moitra, S. D. (1975). Content analysis of references: adjunct or alternative to citation counting? Social Studies of Science, 5(4), 423-441.

Das, B., & Resnick, S. I. (2008). QQ plots, random sets and data from a heavy tailed distribution. Stochastic Models, 24(1), 103-132.

Didegah, F., & Thelwall, M. (2013). Which factors help authors produce the highest impact research? Collaboration, journal and document properties. Journal of Informetrics, 7(4), 861-873.

DORA (2012). San Francisco Declaration on Research Assessment. http://ascb.org/dora/

Eom, Y. H., & Fortunato, S. (2011). Characterizing and modeling citation dynamics. PLoS ONE, 6(9), e24926.

Fairclough, R., & Thelwall, M. (2015). More precise methods for national research citation impact comparisons. Journal of Informetrics, 9(4), 895-906.

Garanina, O. S., & Romanovsky, M. Y. (2015). Citation distribution of individual scientist: Approximations of Stretch Exponential Distribution with Power Law Tails. In Salah, A.A., Y. Tonta, A.A. Akdag Salah, C. Sugimoto, U. Al (Eds.) Proceedings of ISSI 2015. Istanbul, Turkey: Bogaziçi University Printhouse (pp. 272-277).

Gazni, A. (2011). Are the abstracts of high impact articles more readable? Investigating the evidence from top research institutions in the world. Journal of Information Science, 37 (3), 273-281.

Gillespie, C.S. (2013). poweRlaw. https://github.com/csgillespie/poweRlaw

Gillespie, C.S. (2015). Fitting heavy tailed distributions: the poweRlaw package. Journal of Statistical Software, 64(2), 1-16. http://www.jstatsoft.org/v64/i02/paper

Hicks D. (2012). Performance-based university research funding systems. Research Policy 41(2), 251–261.

Hilbe, J. M. (2011). Negative binomial regression (2 ed). Cambridge, UK: Cambridge University Press.

King, D. A. (2004). The scientific impact of nations. Nature, 430(6997), 311-316.

Lawrence, S., Giles, L. C., & Bollacker, K. (1999). Digital libraries and autonomous citation indexing. Computer, 32(6), 67-71.

Levitt, J. & Thelwall, M. (2008). Is multidisciplinary research more highly cited? A macro-level study. Journal of the American Society for Information Science and Technology, 59(12), 1973-1984.

Limpert, E., Stahel, W.A. & Abbt, M. (2001). Lognormal distribution across sciences: Key and clues. Bioscience, 51(5), 341-351.

Low, W. J., Wilson, P. & Thelwall, M. (2015). Stopped sum models for citation data. In: Salah, A.A., Y. Tonta, A.A. Akdag Salah, C. Sugimoto, U. Al (Eds.), Proceedings of ISSI 2015 Istanbul: 15th

International Society of Scientometrics and Informetrics Conference (ISSI 2015), Istanbul: Bogaziçi University Printhouse (pp. 184-194).

MacRoberts, M. H., & MacRoberts, B. R. (1989). Problems of citation analysis: A critical review. Journal of the American Society for Information Science, 40(5), 342-349.

Massey, F.J. (1951). The Kolmogorov-Smirnov test for goodness of fit. Journal of the American Statistical Association, 46(253), 68-78.

Mayr, P., & Walter, A. K. (2007). An exploratory study of Google Scholar. Online information review, 31(6), 814-830.

Merton, R. K. (1973). The sociology of science: Theoretical and empirical investigations. Chicago, IL: University of Chicago press.

Moed, H. F. (2006). Citation analysis in research evaluation. Berlin: Springer Science & Business Media.

Neyman, J. (1939). On a new class of "contagious" distributions, applicable in entomology and bacteriology. The Annals of Mathematical Statistics, 10(1), 35–57. doi:10.1214/aoms/1177732245

Oppenheim, C., & Renn, S. P. (1978). Highly cited old papers and the reasons why they continue to be cited. Journal of the American Society for Information Science, 29(5), 225-231.

Pennock, D. M., Flake, G. W., Lawrence, S., Glover, E. J., & Giles, C. L. (2002). Winners don't take all: Characterizing the competition for links on the web. Proceedings of the national academy of sciences, 99(8), 5207-5211.

Peters, H.P.F. & van Raan, A.F.J. (1994). On determinants of citation scores: A case study in chemical engineering. Journal of the American Society for Information Science, 45 (1), 39-49.

Pettitt, A. N., & Stephens, M. A. (1977). The Kolmogorov-Smirnov goodness-of-fit statistic with discrete and grouped data. Technometrics, 19(2), 205-210.

Radicchi, F., Fortunato, S., & Castellano, C. (2008). Universality of citation distributions: Toward an objective measure of scientific impact. Proceedings of the National Academy of Sciences, 105(45), 17268-17272.

Redner, S. (1998). How popular is your paper? An empirical study of the citation distribution. The European Physical Journal B-Condensed Matter and Complex Systems, 4(2), 131-134.

Rinia, E.J., Van Leeuwen, T.N., & Van Raan, A.F.J. (2002). Impact measures of interdisciplinary research in physics. Scientometrics, 53(2), 241-248.

Schweizer, K. (2010). Judging a journal by the impact factor: Is it appropriate and fair for assessment journals? European Journal of Psychological Assessment. 26(4), 235-237. doi:10.1027/1015-5759/a000031

Seglen, P. O. (1992). The skewness of science. Journal of the American Society for Information Science, 43(9), 628-638.

Smolinsky, L. (2016). Maximum-likelihood estimation for zero inflated distributions. Figshare, https://dx.doi.org/10.6084/m9.figshare.3185113.v1

Thelwall, M. & Fairclough, R. (2015). Geometric journal impact factors correcting for individual highly cited articles. Journal of Informetrics, 9(2), 263–272.

Thelwall, M. & Wilson, P. (2014). Distributions for cited articles from individual subjects and years. Journal of Informetrics, 8(4), 824-839.

Thelwall, M. (2016a). The precision of the arithmetic mean, geometric mean and percentiles for citation data: An experimental simulation modelling approach. Journal of Informetrics, 10(1), 110-123. doi:10.1016/j.joi.2015.12.001

Thelwall, M. (2016b). The discretised lognormal and hooked power law distributions for complete citation data: Best options for modelling and regression. Journal of Informetrics, 10(2), 336-346. doi:10.1016/j.joi.2015.12.007

Thelwall, M. (2016c). Are the discretised lognormal and hooked power law distributions plausible for citation data? Journal of Informetrics, 10(2), 454-470. doi:10.1016/j.joi.2016.03.001

van Raan, A. F., Van Leeuwen, T. N., & Visser, M. S. (2011). Severe language effect in university rankings: particularly Germany and France are wronged in citation-based rankings. Scientometrics, 88(2), 495-498.

van Raan, A. F. (2001). Two-step competition process leads to quasi power-law income distributions: Application to scientific publication and citation distributions. Physica A: Statistical Mechanics and its Applications, 298(3), 530-536.

van Raan, A. F. (2005). Measuring science. In Handbook of quantitative science and technology research (pp. 19-50). Springer Netherlands.

Vuong, Q. H. (1989). Likelihood ratio tests for model selection and non-nested hypotheses. Econometrica: Journal of the Econometric Society, 57(2), 307-333.

Waltman, L., Calero-Medina, C., Kosten, J., Noyons, E., Tijssen, R. J., Eck, N. J., ... & Wouters, P. (2012). The Leiden Ranking 2011/2012: Data collection, indicators, and interpretation. Journal of the American Society for Information Science and Technology, 63(12), 2419-2432.

Waltman, L., van Eck, N. J., van Leeuwen, T. N., Visser, M. S., & van Raan, A. F. (2011). Towards a new crown indicator: An empirical analysis. Scientometrics, 87(3), 467-481.

Waltman, L., & Schreiber, M. (2013). On the calculation of percentile-based bibliometric indicators. Journal of the American Society for Information Science and Technology, 64(2), 372-379.

Wilsdon, J., Allen, L., Belfiore, E., Campbell, P., Curry, S., Hill, S., ... & Kerridge, S. (2015). The metric tide: Report of the independent review of the role of metrics in research assessment and management.
http://www.hefce.ac.uk/pubs/rereports/Year/2015/metrictide/Title,104463,en.html

Wilson, P. (2015). The misuse of the Vuong test for non-nested models to test for zero-inflation. Economics Letters, 127, 51-53.

Zitt, M. & Bassecoulard, E. (1998). Internationalization of scientific journals: a measurement based on publication and citation scope. Scientometrics, 41 (1-2), 255-271.

Zitt, M. (2012). The journal impact factor: Angel, devil, or scapegoat? A comment on JK Vanclay's article 2011. Scientometrics, 92(2), 485-503.

# 10.    Appendix

Videos showing the effect of zero inflation on both models for each subject category are available online in the supplementary material for this article on the journal website and in FigShare https://dx.doi.org/10.6084/m9.figshare.3186997.v1.

The code used to generate the results in the paper is also available online at the same URL: https://dx.doi.org/10.6084/m9.figshare.3186997.v1 (see also: Smolinsky, 2016).